\documentclass[12pt, a4paper]{article}
\usepackage{authblk}

\usepackage[latin1]{inputenc}
\usepackage[T1]{fontenc}
\usepackage{hyperref}

\usepackage{natbib} 
\usepackage{graphicx}

\usepackage{geometry}
\newcommand\third {$3^{\rm rd\,\,\,}$}
\providecommand{\keywords}[1]{Keywords ---  \textit{#1}}
\providecommand{\orcid}[2]{#1 --- \textsc{#2} \\}

\begin{document}
\title{Summary of the \third BINA Workshop}

\author[1]{\textsc{Eugene Semenko}}
\author[2]{\textsc{Manfred Cuntz}}

% % \author{}{}

\affil[1]{National Astronomical Research Institute of Thailand (Public Organization) \\ 260  Moo 4, T. Donkaew,  A. Maerim, Chiangmai, 50180 Thailand}

\affil[2]{Department of Physics, University of Texas at Arlington, Arlington, TX 76019, USA}

\date{}
\maketitle

\begin{abstract}
BINA-3 has been the third workshop of this series involving scientists from India and Belgium aimed at fostering future joint research in the view of cutting-edge observatories and advances in theory.  BINA-3 was held at the Graphic Era Hill University, 22-24 March 2023 at Bhimtal (near Nainital), Uttarakhand, India.  A major event was the inauguration of the International Liquid-Mirror Telescope (ILMT), the first liquid mirror telescope devoted exclusively to astronomy.  BINA-3 provided impressive highlights encompassing topics of both general astrophysics and solar physics.  Research results and future projects have been featured through invited and contributed talks, and poster presentations.
\end{abstract}


%\section{Section -- Level 1 title (Times New Roman, bold, 14 pts)}

\section{Indo-Belgian collaboration in Space and Time}

Without comprehensive international collaborations, it is difficult to imagine sustainable scientific progress in the modern age. In astronomy and astrophysics, such collaborations enabled the operation of observational facilities in the best places on the ground and in space. In big international cooperations like the European Southern Observatory, we can see how the technology exchange and mobility of human resources promote research on all levels, from universities to international institutions. Especially promising collaborations pertain to India, the world's most populous country according to the United Nations \citep{desa}, with exceptionally rapid economic growth.
% \footnote{UN DESA Policy Brief No. 153. The PDF document is available on \url{https://www.un.org/development/desa/dpad/wp-content/uploads/sites/45/PB153.pdf}}

The Belgo-Indian Network for Astronomy and Astrophysics, or BINA, was initialized in 2014 to foster the existing contacts between Indian and Belgian researchers, mostly from the Aryabhatta Research Institute of Observational Sciences (ARIES) and the Royal Observatory of Brussels (ROB), and to expand this collaboration on the nation-wide scale in both countries. The third BINA workshop, which we have the pleasure of summarizing, marks the end of this project. Two previous workshops were held in 2016 in Nainital (India) and 2018 in Brussels (Belgium). We believe that our summary would not be complete without a brief comparison of the third workshop with the two preceding ones. This will help us to better understand BINA's importance and outcome.

The first workshop (BINA-1) took place in Nainital on 15--18 November 2016. According to available statistics \citep{2018BSRSL..87....1D}, 107 astronomers from eight countries participated in the meeting, giving 36 oral talks and presenting 42 posters. Eighty-eight people from twelve partner institutes represented the Indian astronomical community, whereas six Belgian institutions sent ten representatives. The meetings' agenda focused primarily on the instrumentation of the newly commissioned 3.6-m Devastal Optical Telescope (DOT) and on the future of the 4-m International Liquid-Mirror Telescope (ILMT). The scientific talks covered a wide range of subjects, from solar system studies to individual stars, stellar clusters, exoplanets and extragalactic astronomy.

The second BINA workshop (BINA-2) was held two years later, in 2018, in Brussels; it was aimed to further expand the existing collaborations. Despite the significantly smaller number of participants (i.e., 69 registered researchers from seven countries), the conference's scientific programme was rich in oral talks, totalling 44. Furthermore, there were eight poster presentations \citep{2019BSRSL..88....1D}. The scientific programme of the second workshop largely mirrored the agenda of the first meeting, accentuating the scientific application of the Belgo-Indian telescopes. A highly notable aspect of the second workshop's scientific programme was the presence of the review talks.

In terms of participation and the number of oral talks, BINA-3, the final workshop, resembles the previous events, although, fortunately, a significant increase in participation and contributions occurred.  Nearly one hundred fifty scientists from eleven countries participated in BINA-3, with the lion's share from India and Belgium. A total of 37 talks (10: invited, 27: contributory) talks were given in the main programme, and 21 contributory talks were given in the solar physics sessions.  There have been 81 poster presentations; many of those were led by graduate and undergraduate students.

There is significant progress hiding behind the numbers. Since 2016, the Belgo-Indian network has grown to involve new institutes from both partner countries. The members published numerous scientific papers with results obtained on the Belgo-Indian telescopes. Many of these were based on PhD theses pursued within BINA. The content of these proceedings, during 2016--2023, also reveals that many young researchers changed their affiliation, moving to new places and thus expanding the network of research contacts. Progress in instrumentation and scientific collaboration within BINA and with external institutes worldwide gave new impulses to solar and general physics studies. In general, we can count the significantly increased number of telescopes and instruments as the major indicator of progress achieved within the BINA project. The list of available instruments has been highly influential on BINA-3. In the following sections, we briefly summarize its scientific programme.

\section{Observational Techniques and Instrumentation}

Telescopes and their instruments were in the spotlight of all BINA workshops. The ILMT has become the central theme of the current meeting. From a number of oral talks and poster presentations, one could get a comprehensive view of such telescopes' operation principles. It was particularly interesting to find out about the data reduction, calibration and access to the processed images obtained with the ILMT. Numerous results of the first observations with the ILMT, shown mostly in the poster presentations, have demonstrated a wide range of possible scientific applications of zenith telescopes with liquid mirrors. Given the short time that has passed since the beginning of the operation and obtained results, we can confirm that the ILMT has proven its scientific concept and significantly strengthened the observational facilities for the current and future Indo-Belgian projects.

The Indo-Belgian 3.6-m Devastal Optical Telescope (DOT) remains Asia's largest so far fully steerable optical telescope, which has been in operation since 2016. Yet, accurately by the time of BINA-3, a park of Indian telescopes received strengthening with the commissioning of the 2.5-m telescope, which was built by the Advanced Mechanical and Optical Systems (AMOS) in Belgium for the Physical Research Laboratory (PRL) in Ahmedabad and installed at Mt Abu, Rajasthan, India.

The development of new instruments and the upgrade of existing facilities was the central theme of the instrumentation section of the current conference. Notably, by 2028, the TIFR-ARIES Multi-Object Optical to Near-infrared Spectrograph (TA-MOONS) will bring new capabilities useful for the studies of stars in star formation regions, open clusters, and extended sources with DOT. Also, for this telescope, adding the polarimetric mode to the Aries-Devasthal Faint Object Spectrograph \& Camera (ADFOSC), the existing device for observations of faint objects, will enable both linear and circular polarimetry. This new regime is of critical importance to the study of processes in star-forming regions, interacting stellar systems, supernovae, active galactic nuclei, and beyond.

A spectropolarimetric mode might be a case to think of for the creators of the PRL Advanced Radial Velocity Abu Sky Search-2 (PARAS-2), a high-resolution spectrograph at the 2.5-m PRL telescope at Mt Abu. This highly stable device has been developed for precise measurements of radial velocities while providing very high spectral resolution. Due to the geographical location of Mt Abu, PARAS-2 can play a critical role in the continuous monitoring of radial velocities for a wide variety of relatively bright objects; however, with a spectropolarimetric mode being implemented (like HARPSpol at the High Accuracy Radial velocity Planet Searcher (HARPS); \citealt{2011Msngr.143....7P}), PARAS-2 can take its niche in observations of hot magnetic stars, either within Indo-Belgian collaboration or in third-party projects like MOBSTER \citep{2019MNRAS.487..304D}.  (MOBSTER is an acronym for Magnetic OB[A] Stars with TESS: probing their Evolutionary and Rotational properties; it is a collaboration of more than 60 scientists from over the world.)  With the completion of a High-Resolution Spectrograph for the 3.6-m Devastal Optical Telescope (DOT-HRS), the astronomical community of ARIES will possess the ability to independently carry out studies in the fields of asteroseismology and stellar abundances. Again, like in the case of PARAS-2, spectropolarimetry with DOT-HRS is expected to increase the list of potential applications of this device and could further expand the ongoing Nainital-Cape survey of pulsating early-type stars \citep{2000BASI...28..251A, 2001AA...371.1048M,2003MNRAS.344..431J,2006A&A...455..303J, 2009A&A...507.1763J, 2010MNRAS.401.1299J, 2012MNRAS.424.2002J, 2016A&A...590A.116J,2017MNRAS.467..633J, 2022MNRAS.510.5854J}.

The rising number of telescopes in India poses questions about the most adequate time allocation policies and the optimal distribution of observational proposals between existing astronomical facilities. We found that the analysis of the time allocation for the 3.6-m DOT regarding the last six observational cycles, as presented at the workshop, indicated that it was particularly useful and appropriate for all facilities of ARIES --- especially considering that the ILMT has started its operation and the upcoming arrival of the next-generation instruments for the 3.6-m DOT. From our perspective, in addition to the proposed improvements, we would also recommend the organisation of regular (e.g., on a yearly basis) conferences of the telescope's users under the auspices of the Time Allocation Committee (TAC), where the existing and potential applicants would be able to present their proposals or give feedback on the approved or running programmes. Such mini-conferences could be held online, speeding up communication between the TAC and the astronomical community. Naturally, this experience could be applied to other instruments in India and beyond as well.

The theme of small telescopes has been raised in several talks. The Belgium-made High-Efficiency and high-Resolution Mercator Echelle Spectrograph (HERMES), operated at the 1.25-m Mercator telescope in La Palma (Spain), proved its effectiveness in studies of the chemical composition of single and multiple stars. This spectrograph is used for existing bilateral projects. Complimentary opportunities for high-resolution spectroscopy with the 1-m-class telescopes and the perspectives of affordable implementation of adaptive optics on small and moderate-size telescopes have been considered in BINA-3. The interest in these problems highlights the importance of small, properly equipped telescopes for big programmes complementary to missions like the Transiting Exoplanet Survey Satellite (TESS).

\section{Main Programme Session}

BINA provides access to a wide variety of observational facilities located worldwide \citep{2019BSRSL..88....1D}. The observational component mostly determined the agenda of the BINA-3.
Comets, planets, asteroids, and orbital debris were in the third BINA workshop's spotlight, though other topics such as stars, including stellar multiplicity, and compact objects have been discussed.  The selection of objects is largely determined by the areas where optical spectroscopy and photometry are most effective with small and medium-sized telescopes. The exception is the study of planetary atmospheres using the method of stellar occultations. Similar techniques require bigger apertures, and being implemented in a 3--6-m class of telescopes can be very beneficial. The 3.6-m DOT is among those few instruments on the planet which have regularly been used for observation of such events \citep[e.g.,][]{2021ApJ...923L..31S, 2022JAI....1140002S}.

Various instruments available within the Indo-Belgian collaboration promote the comprehensive study of processes occurring in star formation regions and during the ongoing evolution of stars. The efficiency of multi-wavelength observations was demonstrated in the example of the study of the star formation H\,\textsc{ii} region Sh\,2-305. However, this is not a unique case where the Indian telescopes exploring the Universe in optical, radio, and X-ray domains were successfully combined. We cannot pass by the numerous results of the study of massive binary stars, stars with discs and circumstellar envelopes, introduced in the BINA-3 workshop.

Stellar multiplicity runs the golden thread through many talks given in Bhimtal during the workshop. As companions significantly influence stellar lifes at all stages of evolution, proper accounting and evaluation of the companions' properties are crucial. In this regard, work with the catalogues of binary stars or their extensive study within the ongoing or future Indo-Belgian projects must receive high priority. In such programmes, high-resolution optical spectroscopy of binary and multiple stars must take a special place.

Another problem passing through the scientific content of BINA-3 is stellar magnetism. As pointed out in the workshop, magnetic fields are ubiquitous on and beyond the main sequence, with their strengths varying substantially. Magnetic fields are responsible for different kinds of stellar activity and can impact stellar evolution. Besides the theoretical aspects pertaining to the physics of these processes, we would like to attract attention to the lack of observational facilities in the Asian region suitable to direct observations of stellar magnetic fields and processes. The worldwide selection of medium-sized and big telescopes equipped with sensitive spectropolarimetric devices is very limited, and Indian telescopes could fill this gap.

Through the study of chemical composition, one can explore the evolution of individual stars, groups of stars, and the Galaxy at large. The last is the central task of galactic archaeology. Pursuing this task depends on the availability of spectra and proper modelling. Despite the various observational results presented in BINA-3, we find a lack of interactions between the BINA members and groups working, e.g., in the U.S., Sweden or Germany, on the theoretical aspects of abundance analysis. We believe tighter cooperation with the institutes outside of BINA would take the research of stellar abundances to a qualitatively new level.

In contrast to the previous workshops, asteroseismology, a powerful tool for probing stellar interiors and validating stellar parameters, appears underrepresented in BINA-3. (On a lighter note,  a superb cultural show successfully compensated for the lack of ``music of the stars'' in the conference programme.) This fact looks surprising to us as the Belgian groups in Brussels and Leuven are famous for their proficiency in this field.

Apart from galactic archaeology, which deals with the evolution of chemical composition, probing the Galactic structure is another important direction of work within BINA. Even now, after decades of extensive exploration of the Galaxy using different methods, our knowledge of its structure is incomplete. Optical polarimetry helps to reveal the detailed fine structure of dust clouds in the star formation regions or in the areas of young open clusters. Indian astronomers are experienced in this kind of work, and their results, both published \citep[e.g.,][]{2022AJ....164...31U} and presented during BINA-3, deserve special attention. We look forward to further expanding this direction of galactic studies on a new technical level.

\section{Solar Physics Session}

The mainframe of the solar physics programme has been the study of small-scale structure, waves, flares as well as coronal mass ejections (CMEs).  Science opportunities are often directly associated with instruments such as the Extreme Ultraviolet Imager (EUI) onboard of the Solar Orbiter.  The EUI provides a crucial link between the solar surface, on the one hand, and the corona and solar wind, on the other hand, that ultimately shapes the structure and dynamics of the interplanetary medium.  Several contributions focused on wave propagation, including their relevance to small-scale structures of the solar chromosphere, transition region and corona, such as flares, spicules and loop systems.

This kind of research considered both observations and theoretical work, such as ab-initio simulations for standing waves and slow magneto-acoustic waves.  Studies of the outer solar atmosphere also utilized the Interface Region Imaging Spectrograph (IRIS) and the Atmospheric Imaging Assembly (AIA), both onboard of the Solar Dynamics Observatory (SDO).  In alignment with previous studies given in the literature, the potential of spectral lines, including line asymmetries, for the identification of solar atmospheric heating processes has been pointed out and carefully examined.  Clearly, this approach is relevant to both solar physics and studies of solar-type stars of different ages and activity levels; it allows to embed solar studies into a broader context.

Regarding CMEs, a major driver of space weather and geomagnetic stars, attention has been paid the EUropean Heliosphere FORcasting Information Asset (EUHFORIA), which is relevant for MHD modelling and the study of the evolution of CMEs in the heliosphere. In this regard, a pivotal aspect is the study of thermodynamic and magnetic properties of CMEs as well as CME forward-modeling, aimed at predicting CME breakouts as well as CME topologies and magnitudes.  Relevant spectral line features include Fe XIV and Fe XI data, obtained with existing instruments or available in the archive.  Another notable item has been the presentation of long-term variations of solar differential rotation and the solar cycle; the latter still poses a large set of unanswered scientific questions.

\section{Retrospective and Recommendations}

A key element of BINA-3 is the future availability of the ILMT.  The science goals of ILMT include cosmological research such as the statistical determination of key cosmological parameters through surveying quasars and supernovae as well as photometric variability studies of stars, transiting extra-solar planets and various types of transient events.  Another aspect consists in the search for faint extended objects like low-surface brightness and star-forming galaxies.  The pronounced use of ILMT, typically in conjunction with other available facilities, requires the ongoing pursuit of international collaborations; this activity is pivotal for future success.  Another key aspect is the significance of theoretical studies.

Regarding solar physics research, previous work encompasses the study of MHD waves and small-scale transients, with a focus on the solar chromosphere, transition region and corona.  Some of this work made extensive use of the EUI onboard of the Solar Orbiter. The study of outer solar atmosphere fine structure utilized the IRIS and the AIA, both onboard of the SDO. Time-dependent coronal studies, especially CMEs, are of great significance for the Earth, such as the onset of geomagnetic storms and the safety of equipment, including those associated with satellite communication\footnote{See \url{https://www.swpc.noaa.gov} for further information.}.  Further advances in this field are expected to benefit from additional observational studies as well as advances in theory, particularly the interface of those two.  Regarding theoretical work, ongoing and future efforts should continue to focus on 3-D magneto-hydrodynamics studies in conjunction with the adequate inclusion of radiative transfer and statistical phenomena, as well as aspects of chaos theory.

There are other items with the potential for future successful developments. Asteroseismology has been underrepresented in BINA-3.  This is a powerful tool in the context of stellar evolution studies and the validation and improvement of stellar parameters; the latter is also relevant in the context of extrasolar planet investigations.  Further important aspects concern the study of stellar magnetism and activity.  Besides elementary stellar studies, these topics are also of critical importance regarding circumstellar habitability and astrobiology at large \citep[e.g.,][and subsequent work]{2009A&ARv..17..181L}.  Moreover, studies of AGNs and GRBs are cardinal topics beyond solar and stellar physics; they have gained considerable steam within the scientific community.

Processes in the extragalactic objects are characterized by high energy and rich spectra. Among the variety of works presented during BINA-3, studies of active galactic nuclei (AGN) and different transients like gamma-ray bursts (GRB) continue to deserve special attention. The members of BINA have an exhaustive set of instruments available for multi-wavelength observations of these extragalactic sources, yet there is still room for improvement. Considerable advances are attainable both in instrumentation and in techniques of analysis. In the study of intra-night variability of blazars presented in the workshop's programme \citep{2023arXiv230412675A}, we noted the lack of international contributors, although these types of objects are in the spotlight of groups working, e.g., at the 6-m telescope of the Special Astrophysical Observatory, located in the North Caucasus region of Russia \citep{2019MNRAS.482.4322S}. Given the absence of polarimetric devices for observation with the 3.6-m DOT at the moment, such cooperation could open new opportunities. Connections established on the personal level between the member institutions of BINA and observatories operating big telescopes would facilitate future studies in extragalactic astronomy where the aperture matters.

Similarly, we would recommend establishing collaborations with the institutes operating robotic telescopes for the observation of transients. However, a more radical future step might be an expansion of Indian observational facilities towards other continents, especially South America. A small network of medium-sized fully-robotic telescopes could provide easy access to observations and be used for educational purposes.  It would reduce the dependence on astronomical monitoring occurring in South Asia --- in consideration of possible drawbacks due to the regional climates.

Last but not least, in the field of data analysis, the leitmotif now is the use of machine learning (ML) and artificial intelligence (AI). This theme was raised several times during the workshop, but we believe that it could find broader applications in projects related to the classification of light curves and spectra. At the same time, we would recommend researchers using ML and AI in their work not to ignore advances in theory, as without proper constraints and background information, these methods might lead to impractical results, especially if based on small samples.

\subsubsection*{Acknowledgments}
The authors are grateful to the scientific and local organizing committees of BINA-3 for inviting them to summarize the workshop and for further assistance in preparing these proceedings.

\subsubsection*{ORCID identifiers of the authors}
\orcid{0000-0002-1912-1342}{Eugene Semenko}
\orcid{0000-0002-8883-2930}{Manfred Cuntz}

\subsubsection*{Author contributions}
Both authors equally contributed to this publication.

\subsubsection*{Conflicts of interest}
The authors declare no conflict of interest.

\bibliographystyle{apalike}
\bibliography{biblio}

\end{document}